\documentclass[conference]{IEEEtran}
\IEEEoverridecommandlockouts
\usepackage{cite}
\usepackage{amsmath,amssymb,amsfonts}
\usepackage{algorithmic}
\usepackage{graphicx}
\usepackage{textcomp}
\usepackage{xcolor}
\usepackage{caption}
\usepackage{subcaption}
\usepackage[colorlinks]{hyperref}

\makeatletter % changes the catcode of @ to 11
\newcommand{\linebreakand}{%
  \end{@IEEEauthorhalign}
  \hfill\mbox{}\par
  \mbox{}\hfill\begin{@IEEEauthorhalign}
}
\makeatother % changes the catcode of @ back to 12

\DeclareMathOperator{\Tr}{Tr}
\def\BibTeX{{\rm B\kern-.05em{\sc i\kern-.025em b}\kern-.08em
    T\kern-.1667em\lower.7ex\hbox{E}\kern-.125emX}}
\begin{document}

\title{QuantumAnnealing: A Julia Package for Simulating Dynamics of Transverse Field Ising Models}

\author{\IEEEauthorblockN{Zachary Morrell}
\IEEEauthorblockA{\textit{Advanced Network Science Initiative} \\
\textit{Los Alamos National Laboratory}\\
Los Alamos, New Mexico, USA \\
zmorrell@lanl.gov}
\and
\IEEEauthorblockN{Marc Vuffray}
\IEEEauthorblockA{\textit{Advanced Network Science Initiative} \\
\textit{Los Alamos National Laboratory}\\
Los Alamos, New Mexico, USA \\
vuffray@lanl.gov}
\and
\IEEEauthorblockN{Sidhant Misra}
\IEEEauthorblockA{\textit{Advanced Network Science Initiative} \\
\textit{Los Alamos National Laboratory}\\
Los Alamos, New Mexico, USA \\
sidhant@lanl.gov}
\and

\linebreakand

\IEEEauthorblockN{Carleton Coffrin}
\IEEEauthorblockA{\textit{Advanced Network Science Initiative} \\
\textit{Los Alamos National Laboratory}\\
Los Alamos, New Mexico, USA \\
cjc@lanl.gov}
}

\maketitle

\begin{abstract}
Analog Quantum Computers are promising tools for improving performance on applications such as modeling behavior of quantum materials, providing fast heuristic solutions to optimization problems, and simulating quantum systems.  Due to the challenges of simulating dynamic quantum systems, there are relatively few classical tools for modeling the behavior of these devices and verifying their performance.  QuantumAnnealing.jl provides a toolkit for performing simulations of Analog Quantum Computers on classical hardware.
This package includes functionality for simulation of the time evolution of the Transverse Field Ising Model, replicating annealing schedules used by real world annealing hardware, implementing custom annealing schedules, and more.  This allows for rapid prototyping of models expected to display interesting behavior, verification of the performance of quantum devices, and easy comparison against the expected behavior of quantum devices against classical approaches for small systems. The software is provided as open-source and is available through Julia's package registry system.
\end{abstract}


\section{Introduction}
Quantum computing is currently one of the most active areas of research in Physics, Computer Science, and Mathematics.
This has been driven largely by the theoretical demonstration that quantum algorithms could provide an asymptotic speedup for computations such as factoring \cite{Shor1994} and unordered search \cite{Grover1996}.
Since it was proposed in the 1980's \cite{benioff1980computer, Feynman1982}, researchers have been striving to characterize the behavior of quantum devices to understand the limitations of the various proposed architectures.  Quantum computers can roughly be separated into two high level architectures: gate-based \cite{nielsen2010quantum} and analog devices \cite{Albash_2018}.  These architectures are equivalent with respect to which computations they can perform, however there is a polynomial overhead for converting between them \cite{PhysRevLett.99.070502, aharonov2005adiabatic, zobov2007}.  In practice, application performance is impacted by the details of each architecture.  

Simulating the dynamics of quantum systems is widely considered one of the most promising applications of quantum computers \cite{Feynman1982, lloyd1996universal, RevModPhys.86.153}. Various algorithms have been proposed for simulating quantum Hamiltonians on gate-based quantum devices such as product formulas \cite{trotter1959product, Suzuki1976} or Qubitization \cite{Low_2019}, amongst others \cite{Childs2012, PhysRevLett.123.070503, Chen_2021}.
In contrast, analog quantum computers simulate the evolution of quantum Hamiltonians by simply encoding an analogous quantum system into the native Hamiltonian of the hardware and evolving the system from some initial state.  Recent hardware demonstrations have shown promising progress toward simulating quantum systems which are out of reach of classical computers \cite{kim2023evidence, King_2023, semeghini2021probing}.
However, since we are currently in the era of Noisy Intermediate Scale Quantum (NISQ) \cite{Preskill_2018} computation, where quantum hardware computations are distorted by noise and other control errors, one cannot presume a priori that that quantum hardware produces accurate results.
Classical simulation of quantum physics on small systems is the primary method for developing ground truth for validation and verification of an ideal quantum computation.

This work introduces QuantumAnnealing.jl, a software package that simulates the dynamics of analog quantum computers.
This software is motivated by the analog quantum computers from companies such as D-Wave, QuEra, and Pasqal, which are largely used to evolve the dynamics of a time varying Transverse Field Ising Hamiltonian.  This Hamiltonian is commonly used to implement the quantum annealing algorithm \cite{Kadowaki_1998, farhi2000quantum}, from which this software derives its name.  It is intended to provide a fast and easy interface for simulating quantum Hamiltonians and comparing the results against data output from hardware, as well as observing the difference in expected behavior of quantum and classical optimization algorithms on classical Ising models.

\begin{figure*}
    \centering
     \begin{subfigure}[b]{0.32\textwidth}
         \centering
         \includegraphics[width=\textwidth]{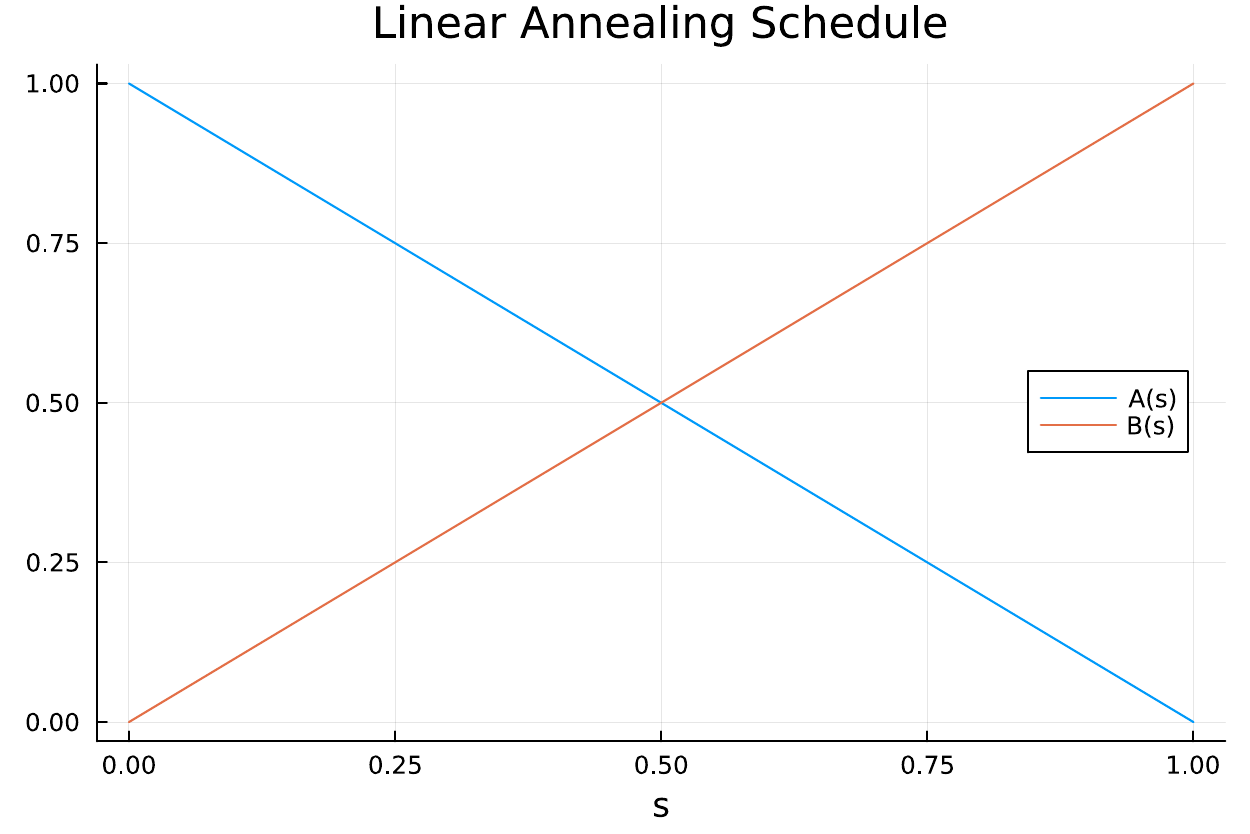}
         \caption{}
         \label{fig:sched_linear}
     \end{subfigure}
    \hfill
     \begin{subfigure}[b]{0.32\textwidth}
         \centering
         \includegraphics[width=\textwidth]{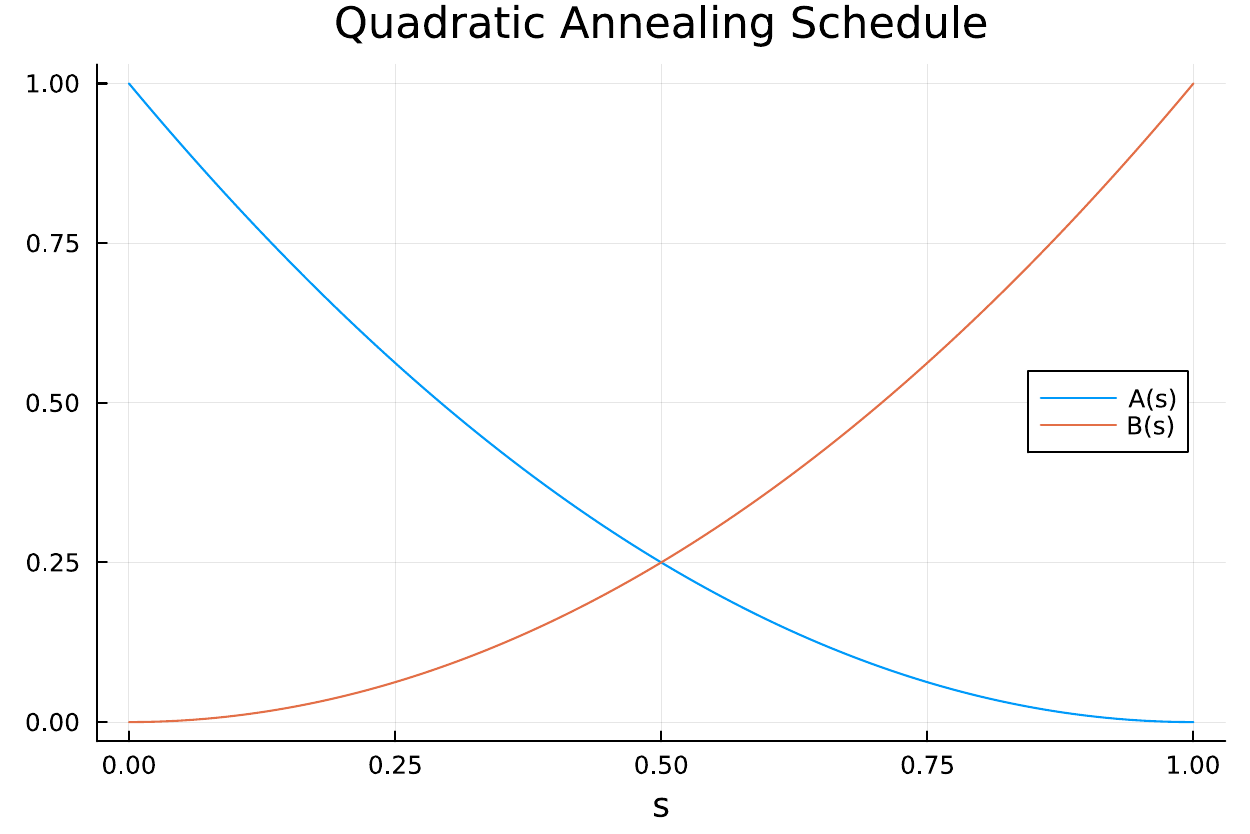}
         \caption{}
         \label{fig:sched_quadratic}
     \end{subfigure}
    \hfill
     \begin{subfigure}[b]{0.32\textwidth}
         \centering
         \includegraphics[width=\textwidth]{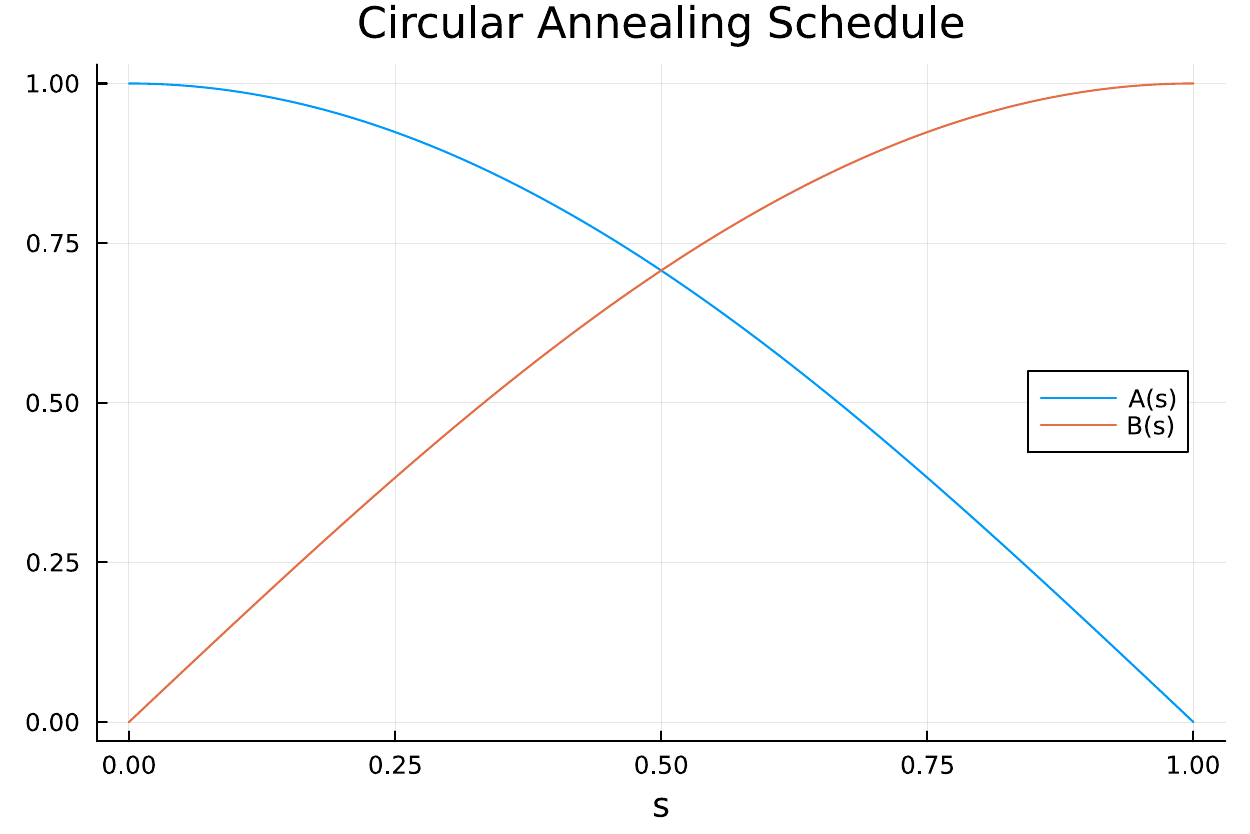}
         \caption{}
         \label{fig:sched_circ}
     \end{subfigure}
    \caption{QuantumAnnealing.jl includes a variety of commonly used annealing schedules.  For a complete list of provided anneailng schedules, see Table \ref{tab:annealing_schedules}.  Figure \ref{fig:sched_linear} shows the linear annealing schedule $A(s) = 1-s$ and $B(s) = s$.  Figure \ref{fig:sched_quadratic} shows the quadratic annealing schedule $A(s) = (s-1)^2$ and $B(s) = s^2$.  Finally, Figure \ref{fig:sched_circ} shows the trigonometric annealing schedule $A(s) = \cos{\left(\frac{\pi}{2} s\right)}$ and $B(s) = \sin{\left(\frac{\pi}{2} s\right)}$. These figures are generated by the associated visualization package, QuantumAnnealingAnalytics.jl, using the \texttt{plot\_annealing\_schedule} function.}
    \label{fig:schedules}
\end{figure*}

While some of the functionality included in QuantumAnnealing.jl is may be achieved by other Hamiltonian simulation packages such as QuTiP \cite{JOHANSSON20121760, JOHANSSON20131234} or HOQST \cite{Chen_2022}, QuantumAnnealing.jl differs with respect to both the underlying solver as well as some added features for convenience, such as support for loading data directly from D-Wave devices, automatically loading annealing schedules from files, and more.  The software also includes functionality provided by commonly used differential equation solving methods by including a wrapper for DifferentialEquations.jl \cite{DifferentialEquations.jl-2017} for use cases where these approaches may be more desirable than the default Magnus expansion based approach.
It should also be noted that QuantumAnnealing.jl focuses on ideal closed-system quantum dynamics and is not currently intended to function as a solver for open-quantum systems. If this functionality is required, one of the other tools mentioned above is likely more suitable.
The current aim QuantumAnnealing.jl is provide a target for ideal quantum computations of models that are ubiquitous in existing analog quantum hardware platforms.
QuantumAnnealing.jl is accompanied by the package QuantumAnnealingAnalytics.jl, which provides additional data visualization capabilities that are use throughout this paper.
Both of these packages are provided as open-source and are available through Julia’s package registry system.\footnote{\href{https://github.com/lanl-ansi/QuantumAnnealing.jl}{https://github.com/lanl-ansi/QuantumAnnealing.jl} \hspace{2cm} \href{https://github.com/lanl-ansi/QuantumAnnealingAnalytics.jl}{https://github.com/lanl-ansi/QuantumAnnealingAnalytics.jl}}

The remainder of the paper is organized as follows.
Section \ref{sec:background} reviews the foundations of quantum simulation and the quantum annealing algorithm specifically. 
Section \ref{sec:methods} reviews the Magnus expansion method for solving ordinary differential equations, which is the primary numerical method used by this work for solving the time varying Schr\"odinger equation.
Section \ref{sec:functionality} provides a brief overview of the primary features of QuantumAnnealing.jl with implementation examples.
The validity and efficacy of the software is shown in Section \ref{sec:validation} through an analysis of various model systems from the literature.
The paper concludes with an outlook of future developments of this software in Sections \ref{sec:future-work} and \ref{sec:conclusion}.  

\section{Background}
\label{sec:background}

\subsection{Quantum Annealing}

\begin{table*}[]
    \centering
    \begin{tabular}{c|c|c}
         Schedule Name &  $A(s)$ & $B(s)$ \\ 
         \hline
         \hline
         \texttt{AS\_LINEAR} & $1-s$ & $s$ \\
         \hline
         \texttt{AS\_CIRCULAR} & $\cos\left(\frac{\pi}{2}s\right)$ & $\sin\left(\frac{\pi}{2}s\right)$\\
         \hline
         \linebreak
         \texttt{AS\_QUADRATIC} & $(1-s)^2$ & $s^2$ \\
         \hline
         \texttt{AS\_DW\_QUADRATIC} & 
         $\begin{array}{lr} \left(13.371976 s^2 -18.453338s + 6.366401 \right) \pi, &
         \text{if } s < 0.69\\ 0, & \text{if } s \geq 0.69\end{array}$
         & $14.55571 (0.85s^2 + 0.15s) \pi $
    \end{tabular}
    \caption{This is a list of the built in annealing schedules in QuantumAnnealing.jl.  The annealing schedules \texttt{AS\_LINEAR}, \texttt{AS\_QUADRATIC}, and \texttt{AS\_CIRCULAR} are common annealing schedules that appear in the literature.  The annealing schedule \texttt{AS\_DW\_QUADRATIC} represents a piece-wise quadratic fit to an annealing schedule associated with the \texttt{DW\_2000Q\_LANL} device from D-Wave.}
    \label{tab:annealing_schedules}
\end{table*}

Quantum annealing is an algorithm based on time varying quantum simulation, which can be used for classical optimization applications.
% Thus far, this quantum algorithm has been studied on analog quantum computing platforms.
The algorithm relies on the adiabatic theorem, which states that if a quantum system is prepared in the ground state of an initial Hamiltonian, under a slow enough evolution the system will remain in the instantaneous ground state.  The quantum annealing procedure is formulated as the evolution of a time varying Hamiltonian
\begin{equation}
    H(s) = A(s) H_{\text{initial}} + B(s) H_{\text{target}}
\end{equation}
where $s \in [0,1]$ is the normalized time parameter defined by $s = \frac{t}{\tau}$,  at time $t$ for total evolution time $\tau$.   $A(s)$ and $B(s)$ are called the annealing schedules of the Hamiltonian.
$H_{\text{initial}}$ and $H_{\text{target}}$ represent the initial Hamiltonian of the system and the Hamiltonian with an unknown ground state which is being optimized, respectively.  The annealing schedules can take many forms, as shown in Figure \ref{fig:schedules}, but for the quantum annealing protocol to work as intended, it should be the case that, $A(0) \gg B(0)$ and $A(1) \ll B(1)$. The dynamics of this system are governed by the Liouville-von Neumann equation,
\begin{equation}
    \frac{d}{ds} \rho(s) = -\frac{i}{\hbar}\tau [H(s), \rho(s)]
\end{equation}
where $\rho$ represents the density matrix of the quantum state, $\hbar$ is the reduced Planck constant, and $i$ is the imaginary unit.  In QuantumAnnealing.jl, we assume that the Hamiltonian is given in natural units, and use $\hbar = 1$.

When implementing quantum annealing in hardware, the initial Hamiltonian is often taken to be a transverse field,
\begin{equation}
    H_{\text{transverse}} = \sum_i \sigma_i^x
\end{equation}
where $\sigma_i^x$ represents the Pauli $X$ matrix operating on qubit $i$.  This Hamiltonian has the easily prepared ground state $|\psi_0\rangle = \bigotimes_i |-\rangle_i$, where $\otimes$ is a Kronecker product.  Often the target Hamiltonian which is used is an Ising Hamiltonian,
\begin{equation}
    H_{\text{ising}} = \sum_{i} h_i \sigma_i^z + \sum_{i,j} J_{ij} \sigma_i^z \sigma_j^z
\end{equation}
where $\sigma_i^z$ represents the Pauli $Z$ matrix operating on qubit $i$.  The coefficients $h_i$ represent the strength of a longitudinal field acting on qubit $i$ and the coefficients $J_{ij}$ represent the coupling strength between qubit $i$ and qubit $j$.
This specific variant of quantum annealing takes the form of a time varying Transverse Field Ising Hamiltonian, 
\begin{equation}
H_\text{TFIM}(s) = A(s) H_{\text{transverse}} + B(s) H_{\text{ising}}
\end{equation}
which is the Hamiltonian that is supported by the most of the current analog quantum computers.

This model is generally used because it is convenient for encoding Quadratic Unconstrained Binary Optimization (QUBO) problems \cite{Lucas_2014} and because this Hamiltonian is easier to realize in a hardware implementation than a more general model.  Since this model is the most ubiquitous, it is currently the focus of QuantumAnnealing.jl.  

\subsection{Example Model}

\begin{figure}
    \centering
    \includegraphics[width=0.48 \textwidth]{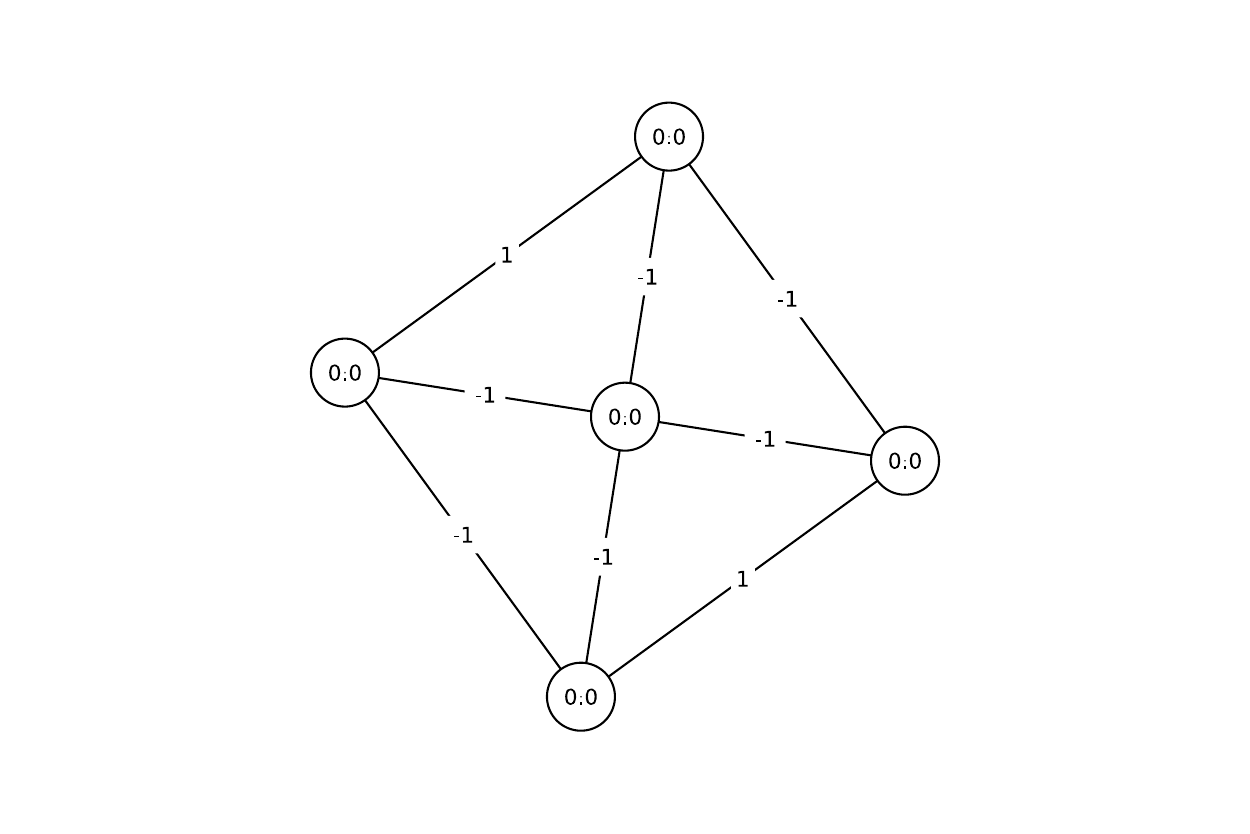}
    \caption{Classical Ising Hamiltonians can be visualized using a graphical structure.  The nodes of the graph represent the local field strengths, $h_i$, and the edges represent the coupling strengths, $J_{ij}$.  For example, the Ising Hamiltonian $H_5$, defined in Equation \eqref{eq:H5} can be plotted graphically using QuantumAnnealingAnalytics.jl with the \texttt{plot\_ising\_model} function. This Ising model was studied extensively in \cite{Matsuda_2009} and will use this model as an example for the purposes of demonstration.
  }
    \label{fig:model_katzgraber}
\end{figure}

\begin{figure}
    \centering
    \includegraphics[width=0.48 \textwidth]{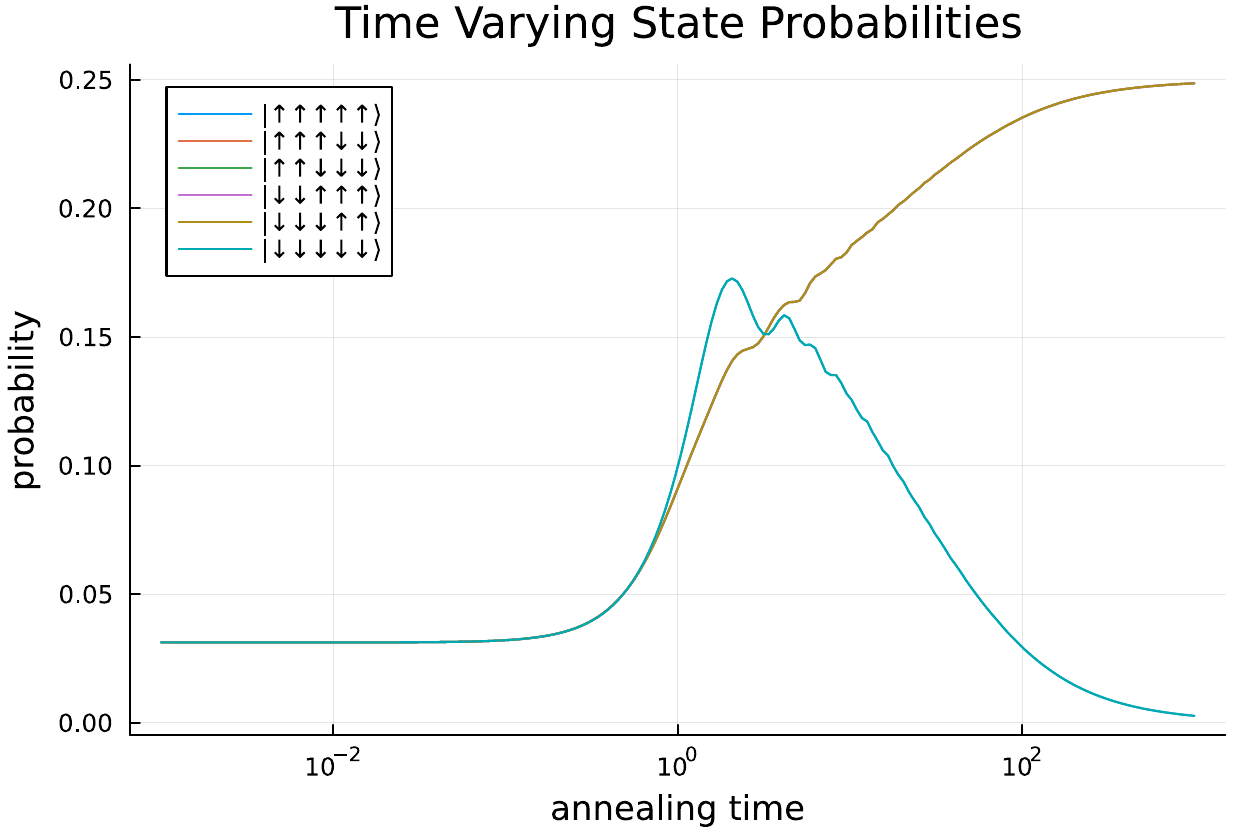}
    \caption{Oftentimes, analog quantum hardware does not have the ability to perform measurements part-way though an evolution ($s < 1$). For this reason, it is often convenient to sweep the total evolution time ($\tau$) to look for signatures that arise from the dynamics of the system being studied.  This protocol is demonstrated above for $H_{5}$ using the \texttt{AS\_CIRCULAR} annealing schedule.  When the simulation time is fast the measurement probabilities of the ground states are uniformly distributed because the dynamics do not have time to propogate through the system.  As the simulation time increases, the states $\lvert\uparrow \uparrow \uparrow \uparrow \uparrow \rangle$ and $\lvert\downarrow\downarrow\downarrow\downarrow\downarrow\rangle$, which have the same measurement probabilities and overlap on the figure, gain a higher measurement probability than the remaining ground states, before being suppressed at long simulation times. This figure can be generated in QuantumAnnealingAnalytics.jl by making use of the \texttt{plot\_varied\_time\_simulations} function.}
    \label{fig:katzgraber_time_varied}
\end{figure}

For the sake of demonstration, we will use the five spin Ising model from \cite{Matsuda_2009}, which has interesting properties in the context of the quantum annealing algorithm.  In this model the $H_{\text{ising}}$ Hamiltonian is defined as,
\begin{equation}
\begin{split}
H_5 = & -\sigma^z_1 \sigma^z_2 - \sigma^z_1 \sigma^z_3 + \sigma^z_1 \sigma^z_4 - \sigma^z_2 \sigma^z_3 \\
      & + \sigma^z_2 \sigma^z_5 - \sigma^z_3 \sigma^z_4 - \sigma^z_3 \sigma^z_5 - \sigma^z_4 \sigma^z_5 
\label{eq:H5}
\end{split}
\end{equation}
which is shown graphically in Figure \ref{fig:model_katzgraber}.  This model was originally studied to compare the ground state measurement statistics of quantum annealing to other classical optimization methods.
It has degenerate ground states with non-uniform, biased measurement probabilities in the adiabatic limit of quantum annealing.
The bias, caused by the driving Hamiltonian, varies wildly, with quite different measurement statistics when the adiabatic condition has and has not been satisfied, as shown in Figure \ref{fig:katzgraber_time_varied}.
This makes it an ideal model for demonstrating the necessity of classical models for verifying the performance of quantum hardware and simulating their expected behavior. For this reason, this model provides a good example use case which we will use throughout this work.


\section{Numerical Methods}
\label{sec:methods}

\subsection{Magnus Expansion}

The preferred method for simulating the evolution of the quantum state $\rho$ in QuantumAnnealing.jl is the \texttt{simulate} function, which applies repeated calls to the Magnus expansion \cite{magnus1954exponential}.  For a pedagogical introduction to the Magnus expansion, see \cite{Blanes_2009, arnal2018general}.  The Magnus expansion is our preferred method because its conservation property ensures unitary time evolution of the quantum state, even when too few iterations are applied for a desired accuracy.  This means that even when too few iterations are applied, approximate workflows can still produce valid quantum states.  Due to this property, the Magnus expansion is also used to simulate time varying Hamiltonians on gate-based quantum computers by generating  time independent unitary operators which can be solved on a gate-based device using either a trotterization or a block encoding approach \cite{An_2022}.

The Magnus expansion solves the system of ordinary differential equations,
\begin{equation}
    \frac{d}{ds}|\psi(s)\rangle = \mathcal{A}(s)|\psi(s)\rangle
\end{equation}
with an initial condition $|\psi(0)\rangle = |\psi_0\rangle$.  The solution is given by,

\begin{equation}
    |\psi(s)\rangle = \exp{\left( \Omega(s) \right)} |\psi_0\rangle
\end{equation}
%where
\begin{equation}
    \Omega(s) = \sum_{k=1}^{\infty} \Omega_k(s)
\end{equation}
with the terms of the sum given by the recursive formula,
\begin{equation}\label{eq:recursive_magnus}
\begin{split}
    \Omega_k(s) &= \sum_{j = 1}^{k-1} \frac{\mathcal{B}_j}{j!} \int_{0}^{s} S_k^{(j)}(s_1) ds_1, \thickspace k > 1  \\
    \Omega_1(s) &= \int_{0}^{s} \mathcal{A}(s_1) ds_1
\end{split}
\end{equation}
where $\mathcal{B}_j$ represents the $j$-th Bernoulli number and the matrices $S_k^{(j)}$ are given by
\begin{equation}
    \begin{split}
        S_{k}^{(1)}(s) &= \left[ \Omega_{k-1}(s), \mathcal{A}(s)\right] \\
        S_{k}^{(j)}(s) &= \sum_{l=1}^{k-j} \left[ \Omega_{l}(s), S_{k-l}^{(j-1)}(s) \right], \thickspace k > j > 1.
    \end{split} \label{eq:magnus-generic-terms}
\end{equation}
The first four terms in the expansion of $\Omega(t)$ can be written explicitly as
\begin{equation}\label{eq:explicit_magnus}
    \begin{split}
        \Omega_1(s) &= \int_{0}^{s} ds_1 \bigl(\mathcal{A}(s_1) \bigr), \\ 
        \Omega_2(s) &= \frac{1}{2}\int_{0}^{s} ds_{1} \int_{0}^{s_{1}} ds_{2} \bigl(\left[ \mathcal{A}(s_1), \mathcal{A}(s_2) \right]\bigr), \\
        \Omega_3(s) &= \frac{1}{6}\int_{0}^{s} ds_{1} \int_{0}^{s_{1}} ds_{2} \int_{0}^{s_{2}} ds_{3} \\
        & \Big( \bigl[ \mathcal{A}(s_1), \left[\mathcal{A}(s_2), \mathcal{A}(s_3) \right] \bigr]\\
        & + \bigl[ \mathcal{A}(s_3), \left[ \mathcal{A}(s_2), \mathcal{A}(s_1) \right] \bigr] \Big),  \\
        \Omega_4(s) &= \frac{1}{12}\int_{0}^{s} ds_{1} \int_{0}^{s_{1}} ds_{2} \int_{0}^{s_{2}} ds_{3} \int_{0}^{s_{3}} ds_{4} \\
        & \biggl( \Bigl[ \bigl[ \left[\mathcal{A}(s_1), \mathcal{A}(s_2) \right], \mathcal{A}(s_3) \bigr], \mathcal{A}(s_4) \Bigr] \\
        & + \Bigl[ \mathcal{A}(s_1), \bigl[ \left[\mathcal{A}(s_2), \mathcal{A}(s_3) \right], \mathcal{A}(s_4) \bigr] \Bigr]\\
        & + \Bigl[ \mathcal{A}(s_1), \bigl[ \mathcal{A}(s_2), \left[ \mathcal{A}(s_3), \mathcal{A}(s_4) \right] \bigr] \Bigr]\\
        & + \Bigl[ \mathcal{A}(s_2), \bigl[ \mathcal{A}(s_3), \left[ \mathcal{A}(s_4), \mathcal{A}(s_1) \right] \bigr] \Bigr]
        \biggr).
    \end{split}
\end{equation}

When applied to a simulating a quantum system, the operator $\mathcal{A}(s) = -i\tau H(s)$ is anti-Hermitian.  This means that when the terms constructing $\Omega(t)$ are truncated by omitting higher order terms to provide an approximation for the full series, this approximation is still unitary, therefore these approximated series are valid quantum operators and the quantum state output by the solver is guaranteed to be properly normalized.  A single step of the Magnus expansion truncated to $\Omega_k$ has error $\epsilon = \mathcal{O}(t^{k+1})$ \cite{sanchez2011new}.  The unitaries constructed by the Magnus expansion are applied to the density operator $\rho(t)$, using the fact that $\rho(t) = \lvert \psi(t) \rangle \langle \psi(t)\rvert$ to obtain the solution,
\begin{equation}
\rho(t) = \exp{\left(\Omega(t) \right)} \rho_0 \exp{\left(\Omega(t) \right)}^{\dagger}
\end{equation}
where $\rho_0 = \rho(0)$.  To apply this process to systems with long annealing times, one must break the evolution into many small time steps with a uniform size,
\begin{equation}
    \Delta t = \frac{\tau}{n_{\text{steps}}}
\end{equation}
where $n_{\text{steps}}$ represents the number of time steps to be used.  This ultimately results in a convergence rate of $\mathcal{O}(\Delta t^{k+1})n_{\text{steps}} = \mathcal{O}(\tau^{k+1}/n_{\text{steps}}^{k})$.  

When $\Delta t$ is too large, the error, $\epsilon$, will also be large.  When $\Delta t$ is too small, $\epsilon$ will also be small, however the computation time will also increase.  To provide a reasonable number of steps for accurate simulation, QuantumAnnealing.jl allows the user to specify two convergence criterion, the maximum element-wise error
\begin{equation}
    E_{\text{max}}(\rho, \hat{\rho}) = \max_{i,j}{(|\rho_{i,j} - \hat{\rho}_{i,j}|)}
    \label{eq:error_max}
\end{equation}
and the mean element-wise error 
\begin{equation}
    E_{\text{mean}}(\rho, \hat{\rho}) = \frac{1}{2^{2n}}\sum_{i,j}{(|\rho_{i,j} - \hat{\rho}_{i,j}|)}
    \label{eq:error_mean}
\end{equation}
where $n$ represents the number of qubits. 
By default QuantumAnnealing.jl will search for a sufficiently large number of steps to achieve the specified accuracy by doubling $n_\text{steps}$ and evaluating the element-wise changes between evaluations. When the solver has cleared the threshold for both convergence criterion, it will return the final density matrix to the user.

In QuantumAnnealing.jl, the recursive formula shown in Equation \ref{eq:recursive_magnus} is implicitly called for simulations where a Magnus expansion of order 5 or higher is desired and can be called explicitly by the user as well. 
%invoking the \texttt{simulate\_magnus\_generic} function.
By default, 
%\texttt{simulate} 
when the user simulates a Hamiltionian's evolution, they will make a call to an  
%\texttt{simulate\_magnus\_optimized} 
optimized function which computes the first four orders of the magnus expansion by using the unwrapped terms shown in Equation \eqref{eq:explicit_magnus}.
To achieve the integration required in each step, the annealing schedule functions, $A(s)$ and $B(s)$ are approximated as quadratic functions through polynomial interpolation, and the integral is computed using the coefficients of the quadratic function.

\section{Functionality}
\label{sec:functionality}

\begin{figure}
    \centering
    \includegraphics[width=0.48\textwidth]{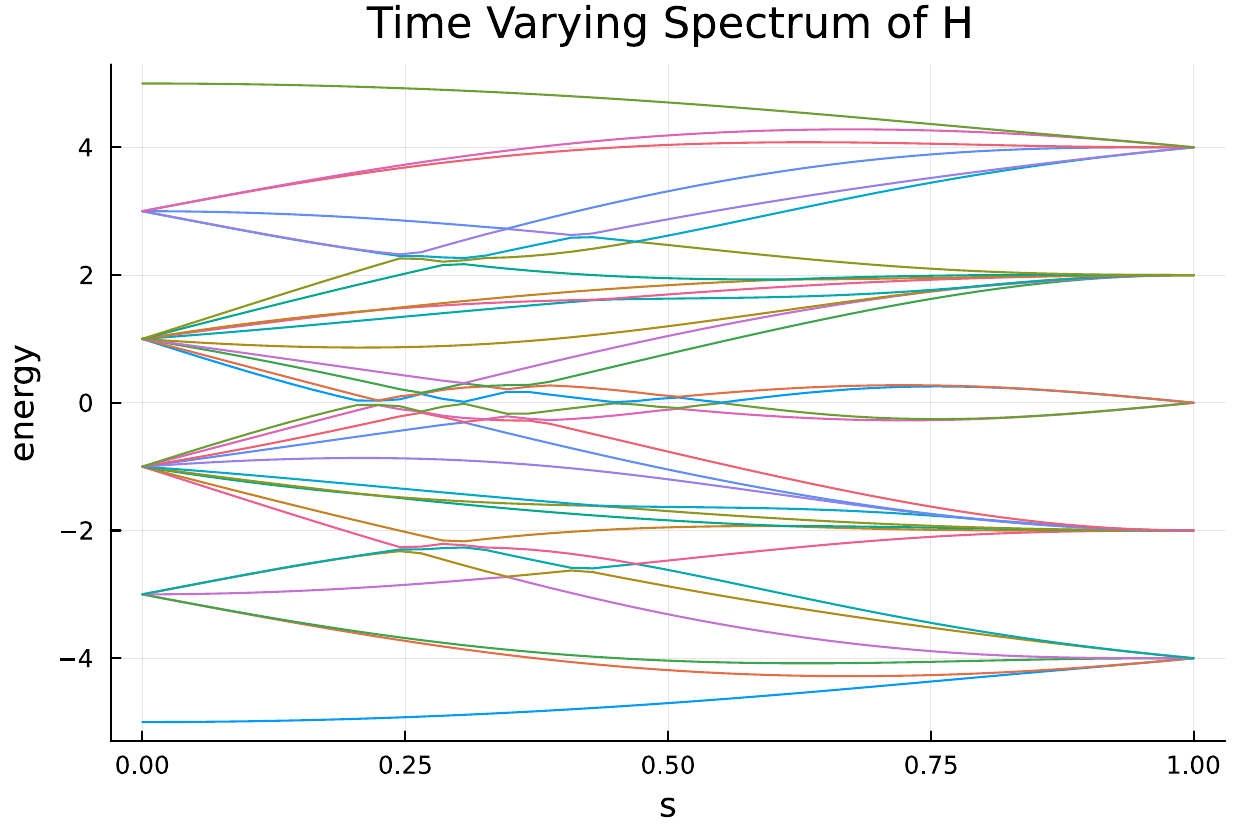}
    \caption{The instantaneous eigenspectrum of a Hamiltonian is important for determining various properties of the dynamics of its evolution, such as the time required to satisfy the adiabatic condition \cite{Albash_2018}. Here we show the eigenspectrum of the Transverse Field Ising Hamiltonian with a uniform transverse field and an Ising component given by $H_{5}$, using the circular annealing schedule, \texttt{AS\_CIRCULAR}, shown in Figure \ref{fig:sched_circ}.}
    \label{fig:energies_katzgraber} 
\end{figure}

This section outlines the general functionality of the QuantumAnnealing.jl software, along with the associated visualization package QuantumAnnealingAnalytics.jl.  QuantumAnnealing.jl is intended to allow simulation of the time evolution of the Transverse Field Ising model in very few lines of code, allowing for simple user implementation of more complicated operations.  Example code snippets demonstrate the workflows for the most common use cases.  First, we will start by showing the code to simulate the time evolution of the Transverse Field Ising model in Equation \eqref{eq:H5} for 100 time units, using the circular annealing schedule \texttt{AS\_CIRCULAR} shown in Figure \ref{fig:sched_linear}.
\begin{verbatim}
import QuantumAnnealing as QA

model_five_spins = Dict(
    (1,2) => -1, (1,3) => -1, (1,4) => 1, 
    (2,3) => -1, (2,5) => 1, (3,4) => -1,
    (3,5) => -1, (4,5) => -1);
annealing_schedule = QA.AS_CIRCULAR
annealing_time = 100
rho = QA.simulate(model_five_spins,
    annealing_time, annealing_schedule)
\end{verbatim}
This example starts by importing the package and defining an alias for brevity (\texttt{QA}), to show which features are being provided by QuantumAnnealing.jl.  Once the package is imported, one can define the $H_\text{target}$ component of the Transverse Field Ising Hamiltonian by using a dictionary. To match Julia convention, the indexing for the spins in the target Hamiltonian begin at 1.  We also select an annealing schedule, in this case \texttt{AS\_CIRCULAR} to define the time varying component of the Hamiltonian, and a total evolution time.  The simulation is performed by calling \texttt{QA.simulate} which applies the Magnus expansion to the initial state for the annealing schedule to return a density matrix, \texttt{rho}.  To define a target Hamiltonian with a local field, one would include one-tuples for the qubits the local field would act on, for example:
\begin{verbatim}
model_five_spins_local_field = Dict(
    (1,2) => -1, (1,3) => -1, (1,4) => 1, 
    (2,3) => -1, (2,5) => 1, (3,4) => -1,
    (3,5) => -1, (4,5) => -1, (1,) => 1,
    (2,) => 1, (3,) => 1, (4,) => 1,
    (5,) => 1);
\end{verbatim}

A useful feature of QuantumAnnealingAnalytics.jl is plotting the instantaneous energy spectrum of the time varying Hamiltonians, as seen in Figure \ref{fig:energies_katzgraber}.  These are useful for determining the energy gaps between the ground state and the various excited states as the anneal progresses.  This allows for estimation of the evolution times required for achiving adiabaticity, as well understanding the effectiveness of  different annealing schedules.

The package also includes support for easy conversion between integer, binary, and spin labels for the various quantum states to match most commonly used labelling conventions.  Data representations are intended to be intuitive, using little-endian ordering of variables in arrays, for example \texttt{binary\_vector = [0, 0, 1]} can be mapped to spin vector using \texttt{binary\_to\_spin(binary\_vector)} to get a vector of spins \texttt{[1, 1, -1]}.  The binary representation can be converted to an integer, following using \texttt{binary\_to\_int} to get the integer value 4.  Lastly the spin vector can be converted to a string in Bra-Ket notation by using \texttt{spin\_to\_braket([1,1,-1])} to get the string "$\lvert \downarrow \uparrow \uparrow \rangle$" and the binary vector can be converted to Bra-Ket notation using \texttt{binary\_to\_braket([0,0,1])} to get the string "$\lvert 100 \rangle$". Note that the Bra-Ket notation uses the big-endian ordering, which is the reverse of the QuantumAnnealing.jl's default.
    
The package includes support for four different annealing schedules.  Three of these schedules can be seen in Figure \ref{fig:schedules}, and all of the built in annealing schedules are defined in Table \ref{tab:annealing_schedules}.  Annealing schedules can be defined to have an initial state of $\bigotimes_i |-\rangle_i$ or $\bigotimes_i |+\rangle_i$, depending on the sign convention of the driving Hamiltonian ($+A(s) H_{\text{driver}}$ and $-A(s) H_{\text{driver}}$ respectively).  This is done to be consistent with the sign convention used by D-Wave.  New annealing schedules can be defined by the user, either by reading in a CSV file with columns of $A(s)$ and $B(s)$ function values which will be interpolated, or by defining the annealing schedules as Julia functions.  For example, one wanted to define a new annealing schedule using cubic functions, one could define it with anonymous functions as follows,
\begin{verbatim}
    AS_CUBIC = QA.AnnealingSchedule(
                    (s) -> s^3, 
                    (s) -> (1-s)^3);
\end{verbatim}

QuantumAnnealing.jl also provides support for the BQPJSON file format \cite{bqpjson}, which allows easy integration with other tools that use this format.
This allows QuantumAnnealing.jl to read the same input and output files that are run on D-Wave's hardware.  From here we can generate histograms of the data, with items sorted either by frequency, hamming weight, or integer value of the states.  Finally, we provide support for time independent field offsets in the $X$ and $Z$ directions, to allow for simulations of experiments like those done in \cite{morrell2023signatures}.

\section{Validation and Performance}
\label{sec:validation}

To validate that the numerical methods implemented in QuantumAnnealing.jl are working properly, two analytically solvable time varying Hamiltonians are compared against the output of the numerical simulations:
\begin{equation}
    H_1(s) = \cos\left(\frac{\pi}{2}s\right) \sigma^x + \sin\left(\frac{\pi}{2}s\right) \sigma^z
\end{equation}
and
\begin{equation}
    H_2(s) = \cos\left(\frac{\pi}{2}s\right) (\sigma^x_1 + \sigma^x_2) + \sin\left(\frac{\pi}{2}s\right) (2 \sigma^z_1 \sigma^z_2).
\end{equation}
The time varying density matrix associated with $H_1$ and $H_2$ can be computed directly.

For compactness, we first define $r = 2\tau$, $\omega_0 = \pi/2$ and $\omega_1 = \sqrt{4 \tau^2 + \pi^2/4}$.  The density matrix, $\rho_{H_1}(t)$ is given by
\begin{equation}
\begin{split}
    \rho_{H_1}(s,\tau) = & \biggl[\bigl(-\left((r^2) + (\omega_0^2)\cos(\omega_1 s)\right) \cos(\omega_0 s)\\
                     & - \omega_0 \omega_1 \sin(\omega_0 s) \sin(\omega_1 s)\bigr)/ (\omega_1^2) \sigma^x\\
                     & -r \omega_0 (1 - \cos(\omega_1 s))/ (\omega_1^2) \sigma^y \\
                     & -\bigl(\left((r^2) + (\omega_0^2)\cos(\omega_1 s)\right) \sin(\omega_0 s) \\
                     & + \omega_0 \omega_1 \cos(\omega_0 s) \sin(\omega_1 s)\bigr)/ (\omega_1^2) \sigma^z\\
                     & + \sigma^I \bigg] / 2
\end{split}
\end{equation}
where $\sigma^I$ is the identity matrix, and $\sigma^x, \sigma^y,$ and $\sigma^z$ are the $X$, $Y$, and $Z$ Pauli matrices respectively.  Similarly, the density matrix for $H_2$ can be computed directly as well.  For the sake of brevity, we will write the elements of the state vector, $\lvert \psi_{H_2}(s,\tau) \rangle$, which can be used to compute the density matrix $\rho_{H_2}(s,\tau) = \lvert \psi_{H_2}(s,\tau)\rangle \langle \psi_{H_2}(s,\tau)\rvert$.  The state vector can be written using the following components
\begin{equation}
\begin{split}
    c_0 = &\cos\left(\frac{\pi}{4} s \sqrt{1+\frac{64 \tau^2}{\pi^2}}\right) \frac{\sqrt{1-\sin\left(\frac{\pi}{2} s\right)}}{2} \\
          &+\left(\frac{8i \tau \sqrt{1-\sin(\frac{\pi}{2} s)}}{\pi} + \sqrt{1+\sin\left(\frac{\pi}{2} s\right)}\right) \\
          & \cdot \left(\frac{\sin\left(\frac{\pi}{4} s \sqrt{1+\frac{64 \tau^2}{\pi^2}}\right) }{2 \sqrt{1+\frac{64 \tau^2}{\pi^2}}}\right)
\end{split}
\end{equation}
%To verify that there were no typos in converting to LaTeX, here is the coded expression from QA.jl's test files
%    c_0 = 1/2*cos(π/4*s*sqrt(1+64*t^2/π^2))*sqrt(1-sin(π/2*s)) +
%        (8im*t*sqrt(1-sin(π/2*s))/π + sqrt(1+sin(π/2*s)))*sin(π/4*s*sqrt(1+64*t^2/π^2)) / (2*sqrt(1+64*t^2/π^2))
and
\begin{equation}
\begin{split}
    c_1 = &-\Biggl[\cos\left(\frac{\pi}{4} s \sqrt{1+\frac{64 \tau^2}{\pi^2}}\right) \left(1+\sin\left(\frac{\pi}{2} s\right)\right)  \\
          &+ \Biggl(\left(-\cos\left(\frac{\pi}{2} s\right) + \frac{8i \tau \left(1+\sin\left(\frac{\pi}{2} s\right)\right)}{\pi}\right) \\
          &\cdot \sin\left(\frac{\pi}{4} s \sqrt{1+\frac{64 \tau^2}{\pi^2}}\right)\Biggr)\frac{1}{\sqrt{1+\frac{64 \tau^2}{\pi^2}}} \Biggr]\\
          & \div \left(2 \sqrt{1+\sin\left(\frac{\pi}{2} s\right)}\right)
\end{split}
\end{equation}
%To verify that there were no typos in converting to LaTeX, here is the coded expression from QA.jl's test files
%    c_1 = -(cos(π/4*s*sqrt(1+64*t^2/π^2))*(1+sin(π/2*s)) + 
%        ((-cos(π/2*s) + 8im*t*(1+sin(π/2*s))/π)*(sin(π/4*s*sqrt(1+64*t^2/π^2))))/(sqrt(1+64*t^2/π^2))
%        )/(2*sqrt(1+sin(π/2*s)))
to obtain the state vector
\begin{equation}
\lvert \psi_{H_2}(s,\tau)\rangle = 
\begin{bmatrix}
    c_0 \\
    c_1 \\
    c_1 \\
    c_0 
\end{bmatrix}.
\end{equation}
By testing that the entries of the density matrices output by the  Magnus expansion solvers are approximately the same as the entries in $\rho_1(s,\tau)$ and $\rho_2(s,\tau)$, we are able to verify that the implementations of the Magnus expansion are correct.

As an example of general performance, we return to the example Hamiltonian, $H_5$.  To evaluate the emperical rate of convergence of the software, we need a reliable metric for the accuracy of the computation.  A commonly used metric for quantum mechanical systems is the trace distance, defined as
\begin{equation}
    D(\rho_1, \rho_2) = \frac{1}{2} \Tr\left[ \sqrt{(\rho_1 - \rho_2)^{\dagger}(\rho_1 - \rho_2)} \right]
    \label{eq:trace_distance}
\end{equation}
which we will use here.

To serve as a baseline for determining error estimates for a representative use case, we perform a simulation of the evolution of the five qubit model $H_5$, defined in Equation \eqref{eq:H5}, evolving for 100 units of time using 100,000 iterations of the Magnus expansion to serve as a surrogate for the exact density matrix which cannot be found analytically.  We will refer to the state output by this simulation as $\rho_{H_5}(100)$.

\begin{figure}
    \centering
    \includegraphics[width=0.95 \linewidth]{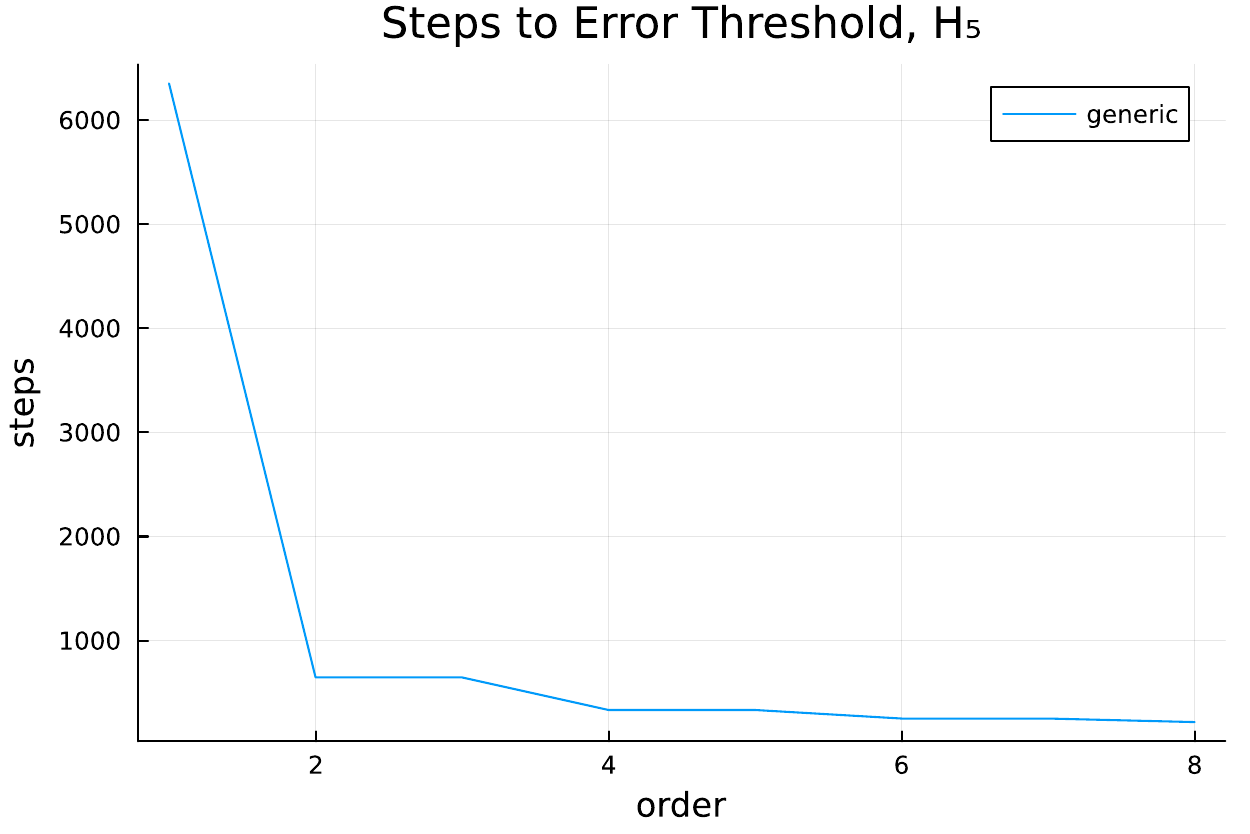}
    \caption{The number of  Magnus expansion steps required for the density matrix to reach a trace distance of $10^{-6}$ from a ground truth density matrix  decreases as the order of the Magnus expansion increases.  This instance shows the number of steps required to reach a trace distance of $10^{-6}$ from the density matrix for $\rho_{H_5}(100)$ generated through the application of 100,000  Magnus expansion steps. Odd orders do not gain a performance improvement due to the structure of the Transverse Field Ising Hamiltonian.}
    \label{fig:steps_to_error_threshold}
\end{figure}

\begin{figure}
    \centering
    \includegraphics[width=0.95 \linewidth]{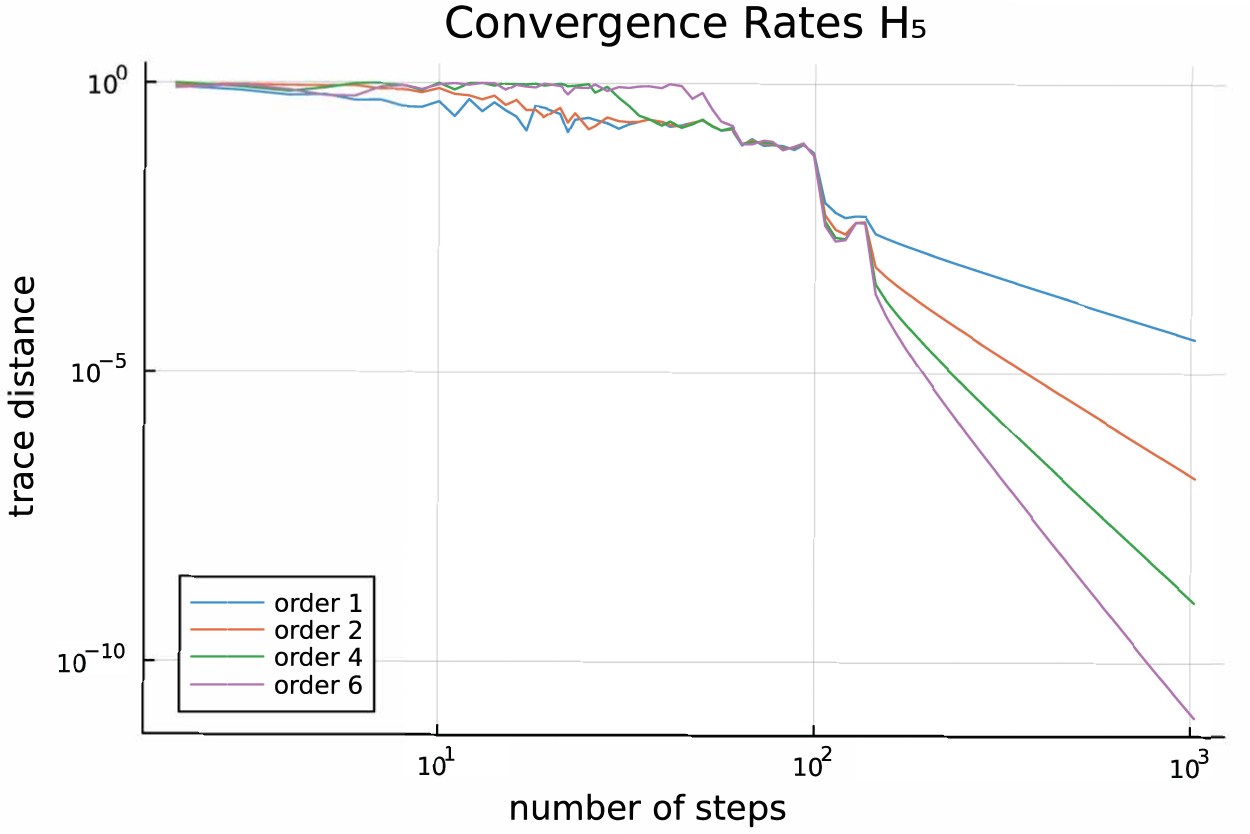}
    \caption{As the number of repeated applications of the Magnus expansion increases ($n_{\text{steps}}$) above a certain convergence threshold which is Hamiltonian dependent, higher order implementations approach convergence more rapidly.  This example shows the convergence rate for $H_5$.  Similar plots are shown for $H_1$ and $H_2$ in Appendix A.  Due to the structure of the Transverse Field Ising Hamiltonian, order 3 and 5 expansions do not gain a convergence improvement over orders 2 and 4 respectively, as demonstrated in Figure \ref{fig:steps_to_error_threshold}.  For clarity, we only show the convergence rates for orders 1, 2, 4, and 6.}
    \label{fig:error_scaling_by_steps}
\end{figure}

To observe the convergence rate of the various orders of the Magnus expansion solver, we first perform the 100 time unit evolution of $H_5$ until a trace distance of $10^{-6}$ is reached and we perform the 100 time unit evolution of $H_5$ using a varying number of iterations of the Magnus expansion.  The results of these experiments can be seen in Figure \ref{fig:steps_to_error_threshold} and Figure \ref{fig:error_scaling_by_steps} respectively.  We observe that due to the structure of the Transverse Field Ising Hamiltonian, that there is generally not a performance gain for odd orders greater than 1.  This motivates the use of the fourth order Magnus expansion as the default solver. Similar validation experiments for the $H_1$ and $H_2$ models are discussed in the additional figures in the appendix.

\begin{figure}
    \centering
    \includegraphics[width=0.95 \linewidth]{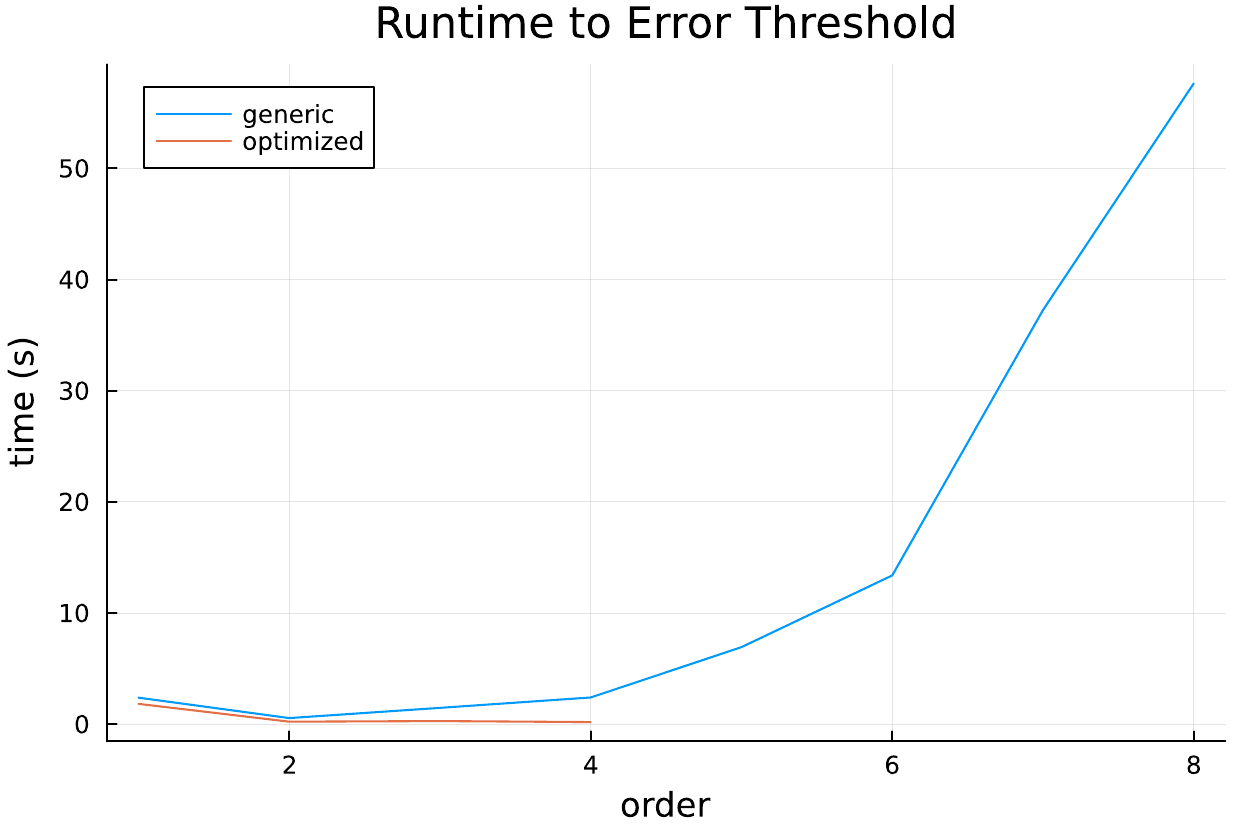}
    \caption{A comparison of the runtime of different order Magnus expansions to reach an error threshold of a trace distance of $10^{-6}$ from $\rho_{H_5}(100)$. There is a clear trade off in Magnus order and run time due to the complexity of implementing higher order Magnus expansions. An optimized implementation of 4th order Magnus is provided as the default implementation in QuntumAnnealing.jl to minimize the amount of time needed to compute accurate simulations. }
    \label{fig:runtime_to_error_threshold}
\end{figure}

Although the increasing Magnus order reduces the amount of steps required to reach a specific accuracy.
It does not necessarily reduce the amount of classical compute time required, because as the Magnus order increases so does the complexity of computing each  Magnus expansion step (there is a combinatorial increase in terms per Equation \eqref{eq:magnus-generic-terms}).
Figure \ref{fig:runtime_to_error_threshold} illustrates the trade-off in runtimes for specific Magnus orders and identifies a dramatic increase in compute time above order 4.
Noting that 4th-order Magnus appears to provide the best trade-off in accuracy and speed, QuantumAnnealing.jl includes a highly optimized software implementation of this order.
The performance benefits of this optimized implementation are also shown in Figure \ref{fig:runtime_to_error_threshold}.

\section{Future Work}
\label{sec:future-work}

There are many potential features which may be added in the future. One of the most immediately aspects which may be beneficial is adding support for more generic Hamiltonians.  This could be in the form of of allowing for simulation of arbitrary weighting of the driving Hamiltonian, or allowing for more Hamiltonian terms to be introduced with $\sigma_i^y$ or $\sigma_i^x \sigma_j^x$ and $\sigma_i^y \sigma_j^y$ terms.  This could allow for the study of the dynamics of more general classes of Hamiltonians such as the Kitaev model or the Heisenberg model.

Another potential area of remaining work is incorporating support for acceleration with GPUs or distributed computing resources for improved performance.  This would potentially allow for more rapid simulation of high time simulations or allow for larger systems to be simulated by taking advantage of shared memory.  Currently QuantumAnnealing.jl is capable of simulating systems on the order of at most 16 qubits, depending on the amount of classical computer memory. Some packages have been able to surpass this number by taking advantage of symmetry arguments \cite{bloqade2023quera} and similar methods may prove applicable for this package as well.  For system sizes which are currently supported, another possible improvement would be the incorporation of adaptive time-stepping to minimize the number of repeated matrix exponentiations required to simulate the system.

It has been shown that a combination of control errors and open system effects best describe the behavior of D-Wave quantum annealing devices \cite{morrell2023signatures}.  QuantumAnnealing.jl currently provides support for random control errors in the form of constant time independent Hamiltonian offsets, but open quantum simulation is not currently supported.  In the future, we may include some functionality for calling methods from OpenQuantumTools.jl \cite{Chen_2022} to allow for a more direct pipeline to switch from closed to open system simulation.

Lastly, there are many relevant device specific utility functions which could be implemented.  For example, D-Wave supports several solver parameters, e.g. modifying the time evolution by changing the rate of change of the normalized time $s$ through their \texttt{annealing\_schedule} parameter.  While this functionality is currently supported by QuantumAnnealing.jl, parameters such as their \texttt{anneal\_offset} are not currently implemented.

\section{Conclusion}
\label{sec:conclusion}

This work explores the challenges of simulating time varying Hamiltonians arising in the context of quantum annealing and puts forth QuantumAnnealing.jl and its associated plotting package QuantumAnnealingAnalytics.jl, as tools to simulate these dynamical systems in a closed-system setting with high accuracy.  This work has shown that QuantumAnnealing.jl serves as a useful tool for prototyping models to identify interesting characteristics of their dynamics, generating commonly used figures found in the literature, and evaluating the performance of quantum annealing hardware.  The use of a Magnus expansion solver for time varying Hamiltonian dynamics will also prove useful in the future for evaluating the performance of gate-based Hamiltonian simulators as well.

\section{Acknowledgements}

Research presented in this paper was supported by the Laboratory Directed Research and Development program of Los Alamos National Laboratory under project numbers 20240032DR, 20230546DR and 20210114ER. This research used resources provided by the Los Alamos National Laboratory Institutional Computing Program, which is supported by the U.S. Department of Energy National Nuclear Security Administration under Contract No. 89233218CNA000001.

\appendix[A. Convergence to Analytic Solutions]
\label{sec:convergence}

To verify that our implementations of the Magnus expansion are correct, we show the convergence to $\rho_{H_1}(100)$ and $\rho_{H_2}(100)$.  Since these models have analytical solutions, we are able to precisely verify the performance of these models, relative to machine epsilon for 64 bit floating point numbers.  We consider three different distance metrics: The trace distance defined in Equation \eqref{eq:trace_distance}, the maximum element-wise error defined in Equation \eqref{eq:error_max},  and the mean element-wise error defined in Equation \eqref{eq:error_mean} 

\begin{figure*}
    \centering
    \begin{subfigure}[b]{0.48\textwidth}
        \centering
        \includegraphics[width=0.95 \linewidth]{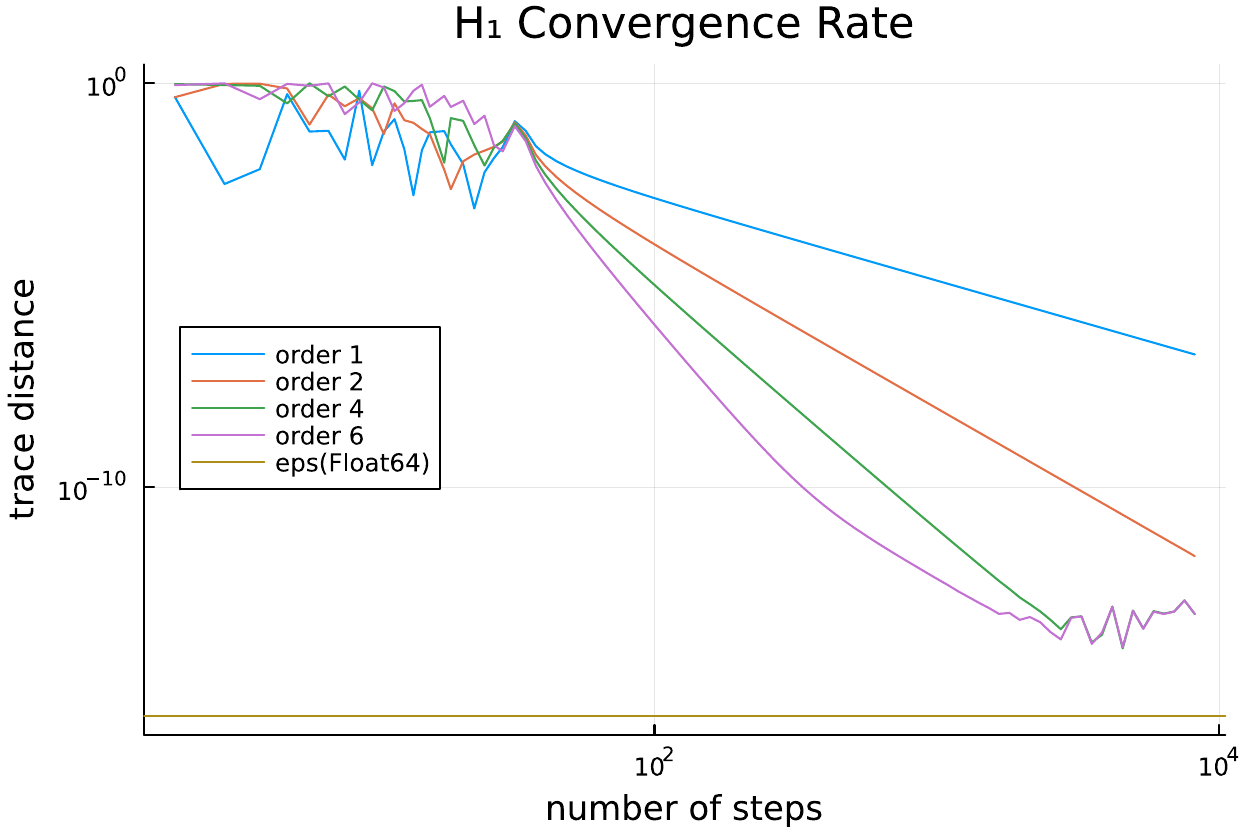}
        \caption{$H_1$}
        \label{fig:error_scaling_by_steps_H1}
    \end{subfigure}
    \hfill
    \begin{subfigure}[b]{0.48\textwidth}
        \centering
        \includegraphics[width=0.95 \linewidth]{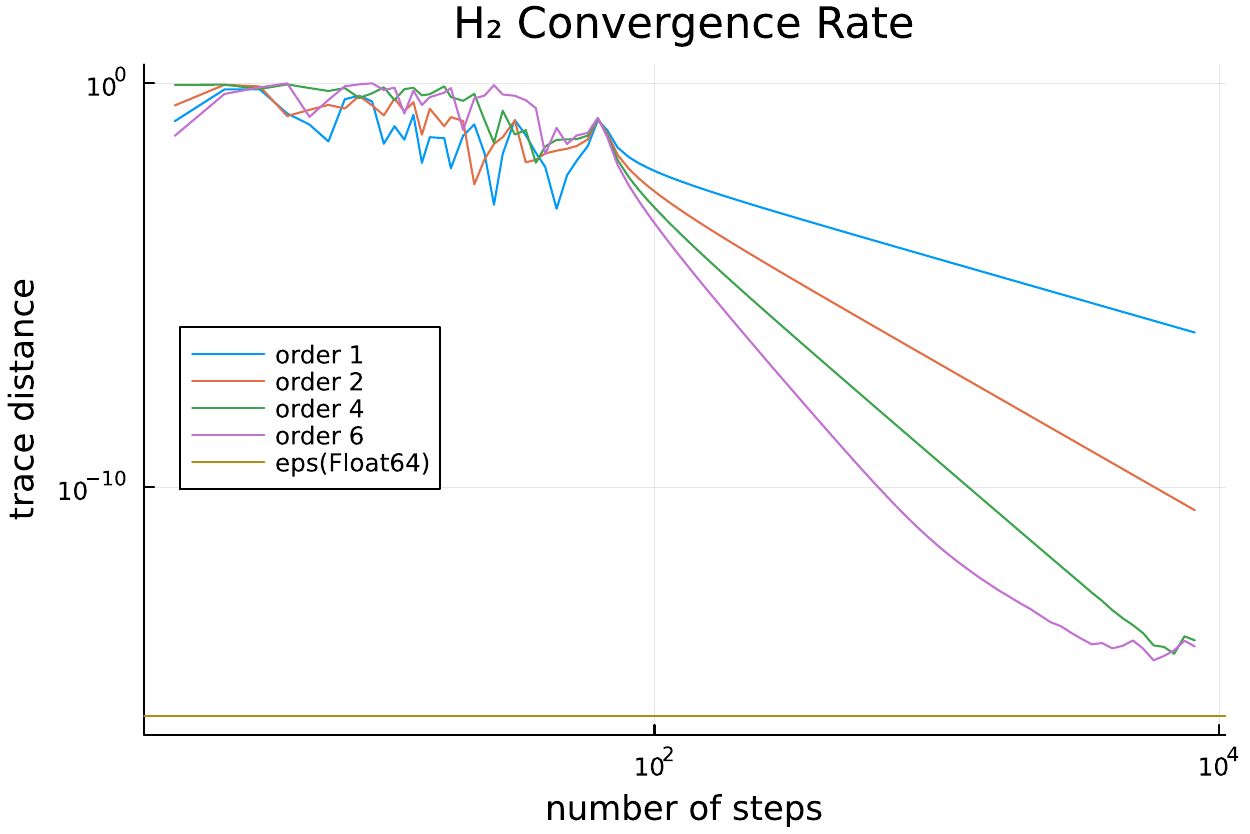}
        \caption{$H_2$}
        \label{fig:error_scaling_by_steps_H2}
    \end{subfigure}
    \caption{The convergence of our implementation can be observed by comparing the trace distances of the analytically solved solutions $\rho_{H_1}(100)$ and $\rho_{H_2}(100)$ for $H_1$ and $H_2$ respectively at 100 time-steps.  We see that we are able to achieve accuracy on the order of $10^{-13}$.  Machine epsilon for 64 bit floating point numbers is shown by the horizontal line to serve as a lower possible bound for accuracy.}
    \label{fig:error_scaling_by_steps_analytic}
\end{figure*}

\begin{figure*}
    \centering
    \begin{subfigure}[b]{0.48\textwidth}
        \centering
        \includegraphics[width=0.95 \linewidth]{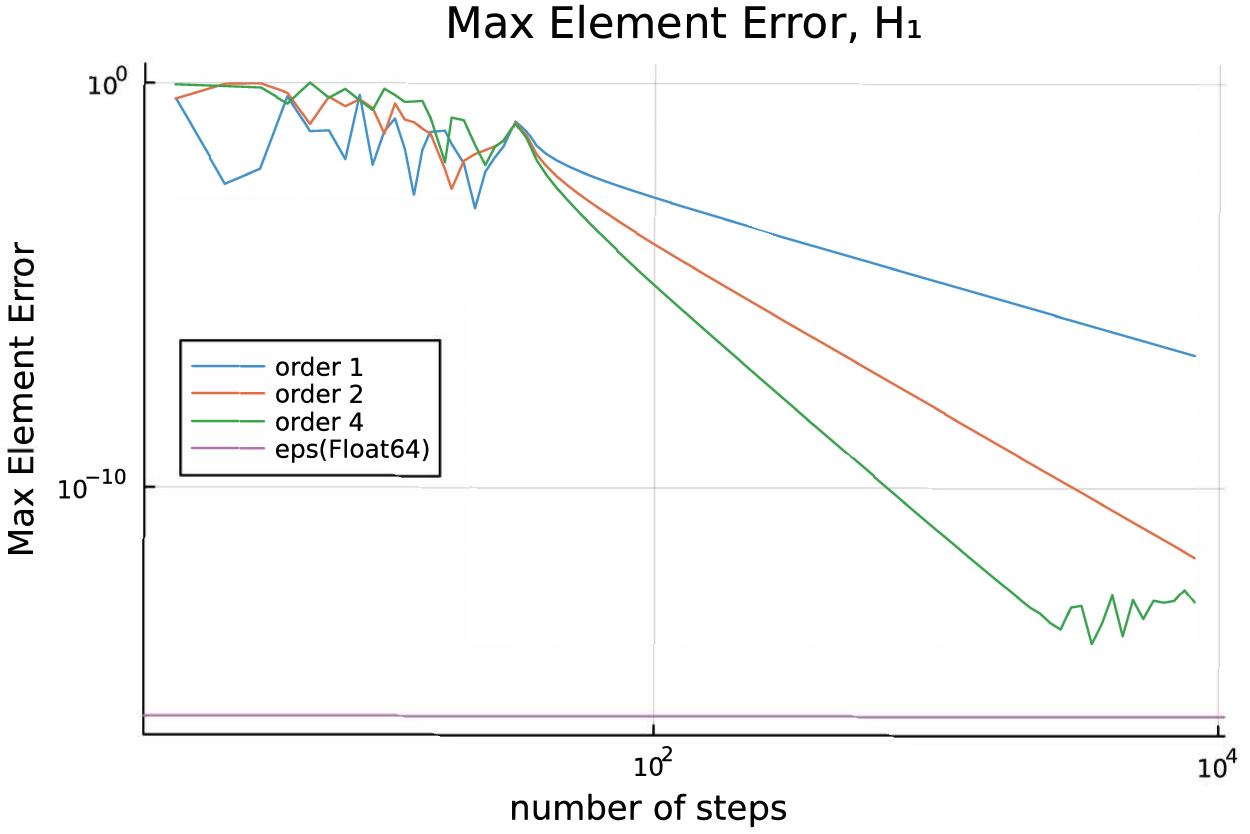}
        \caption{$H_1$}
        \label{fig:error_max_scaling_by_steps_H1}
    \end{subfigure}
    \hfill
    \begin{subfigure}[b]{0.48\textwidth}
        \centering
        \includegraphics[width=0.95 \linewidth]{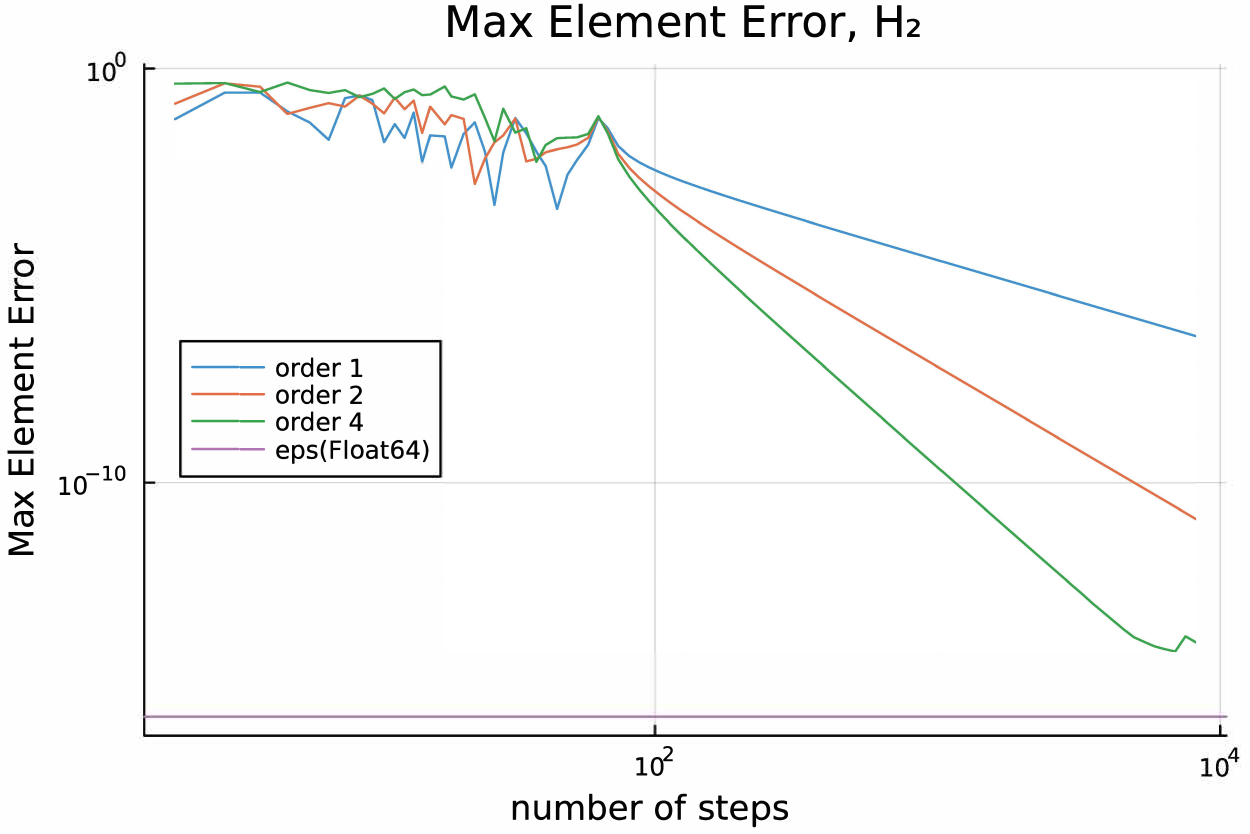}
        \caption{$H_2$}
        \label{fig:error_max_scaling_by_steps_H2}
    \end{subfigure}
    \caption{The maximum element-wise error for our solver plateaus in performance after reaching a value of $10^{-14}$}
    \label{fig:error_max_scaling_by_steps_analytic}
\end{figure*}

\begin{figure*}
    \centering
    \begin{subfigure}[b]{0.48\textwidth}
        \centering
        \includegraphics[width=0.95 \linewidth]{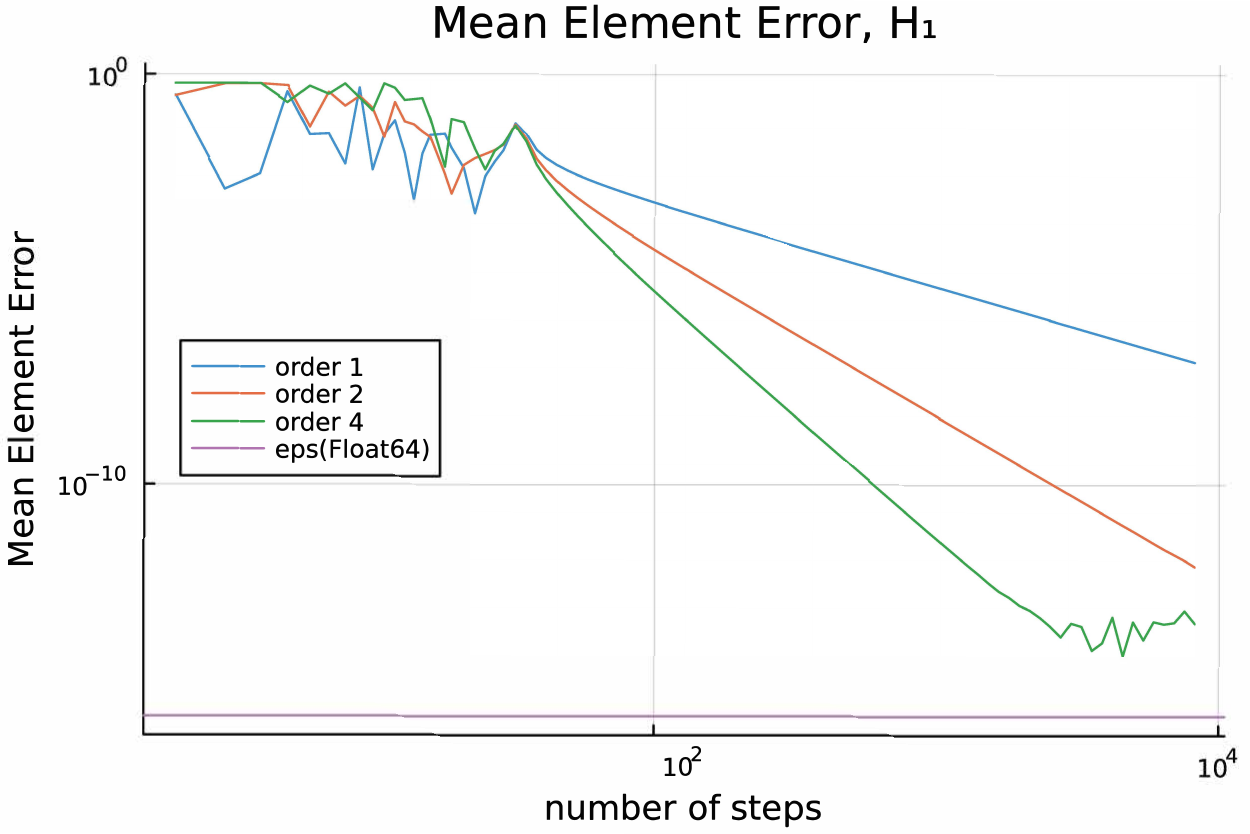}
        \caption{$H_1$}
        \label{fig:error_mean_scaling_by_steps_H1}
    \end{subfigure}
    \hfill
    \begin{subfigure}[b]{0.48\textwidth}
        \centering
        \includegraphics[width=0.95 \linewidth]{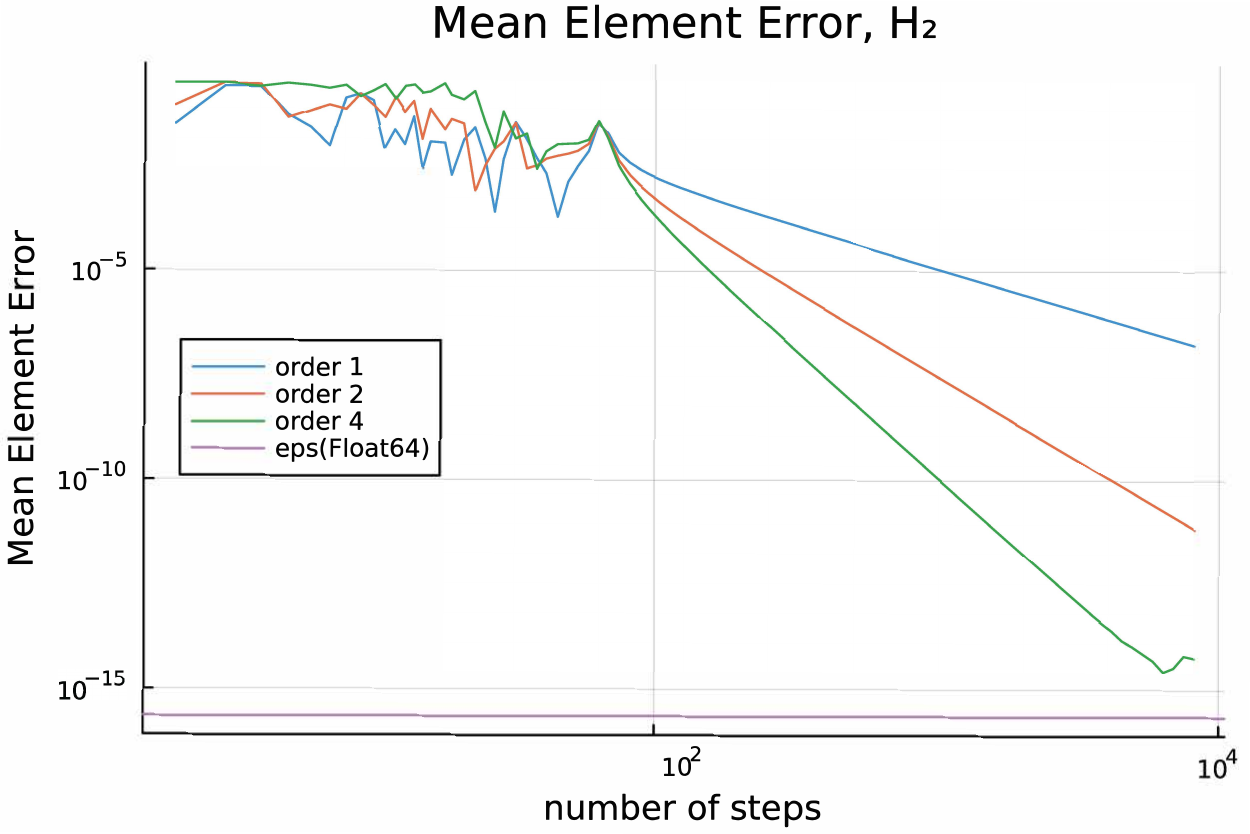}
        \caption{$H_2$}
        \label{fig:error_mean_scaling_by_steps_H2}
    \end{subfigure}
    \caption{The mean element-wise error stops improving after reaching a value on the order of $10^{-14}$.}
    \label{fig:error_mean_scaling_by_steps_analytic}
\end{figure*}

The performance of our solvers on these distance metrics are shown in Figures \ref{fig:error_scaling_by_steps_analytic}, \ref{fig:error_max_scaling_by_steps_analytic}, and \ref{fig:error_mean_scaling_by_steps_analytic}.  We find that the tool is capable of achieving accuracies with a trace distance on the order of $10^{-13}$, and both maximum element-wise error and mean element-wise error on the order of $10^{-14}$ for both the time dynamics of both Hamlitonians.  We notice a slight loss of performance for the sixth order solver after obtaining a trace distance on the order of $10^{-10}$, further motivating our use of the optimized fourth order solver.

%%%Adding manual final page column balancing
\IEEEtriggeratref{26}
\bibliographystyle{ieeetr}
\bibliography{Bibliography}    

\begin{thebibliography}{10}

\bibitem{Shor1994}
P.~Shor, ``Algorithms for quantum computation: discrete logarithms and
  factoring,'' in {\em Proceedings 35th Annual Symposium on Foundations of
  Computer Science}, pp.~124--134, 1994.

\bibitem{Grover1996}
L.~K. Grover, ``A fast quantum mechanical algorithm for database search,'' in
  {\em Proceedings of the Twenty-Eighth Annual ACM Symposium on Theory of
  Computing}, STOC '96, (New York, NY, USA), p.~212–219, Association for
  Computing Machinery, 1996.

\bibitem{benioff1980computer}
P.~Benioff, ``The computer as a physical system: A microscopic quantum
  mechanical hamiltonian model of computers as represented by turing
  machines,'' {\em Journal of statistical physics}, vol.~22, pp.~563--591,
  1980.

\bibitem{Feynman1982}
R.~P. Feynman, ``Simulating physics with computers,'' {\em International
  Journal of Theoretical Physics}, vol.~21, pp.~467--488, Jun 1982.

\bibitem{nielsen2010quantum}
M.~A. Nielsen and I.~L. Chuang, {\em Quantum computation and quantum
  information}.
\newblock Cambridge university press, 2010.

\bibitem{Albash_2018}
T.~Albash and D.~A. Lidar, ``Adiabatic quantum computation,'' {\em Reviews of
  Modern Physics}, vol.~90, Jan. 2018.

\bibitem{PhysRevLett.99.070502}
A.~Mizel, D.~A. Lidar, and M.~Mitchell, ``Simple proof of equivalence between
  adiabatic quantum computation and the circuit model,'' {\em Phys. Rev.
  Lett.}, vol.~99, p.~070502, Aug 2007.

\bibitem{aharonov2005adiabatic}
D.~Aharonov, W.~van Dam, J.~Kempe, Z.~Landau, S.~Lloyd, and O.~Regev,
  ``Adiabatic quantum computation is equivalent to standard quantum
  computation,'' 2005.

\bibitem{zobov2007}
V.~E. Zobov and A.~S. Ermilov, ``Realizations of standard quantum computational
  circuits by adiabatic evolution,'' {\em Theoretical and Mathematical
  Physics}, vol.~150, no.~3, pp.~393--402, 2007.

\bibitem{lloyd1996universal}
S.~Lloyd, ``Universal quantum simulators,'' {\em Science}, vol.~273, no.~5278,
  pp.~1073--1078, 1996.

\bibitem{RevModPhys.86.153}
I.~M. Georgescu, S.~Ashhab, and F.~Nori, ``Quantum simulation,'' {\em Rev. Mod.
  Phys.}, vol.~86, pp.~153--185, Mar 2014.

\bibitem{trotter1959product}
H.~F. Trotter, ``On the product of semi-groups of operators,'' {\em Proceedings
  of the American Mathematical Society}, vol.~10, no.~4, pp.~545--551, 1959.

\bibitem{Suzuki1976}
M.~Suzuki, ``Generalized trotter's formula and systematic approximants of
  exponential operators and inner derivations with applications to many-body
  problems,'' {\em Communications in Mathematical Physics}, vol.~51,
  pp.~183--190, Jun 1976.

\bibitem{Low_2019}
G.~H. Low and I.~L. Chuang, ``Hamiltonian simulation by qubitization,'' {\em
  Quantum}, vol.~3, p.~163, jul 2019.

\bibitem{Childs2012}
A.~M. Childs and N.~Wiebe, ``Hamiltonian simulation using linear combinations
  of unitary operations,'' {\em arXiv preprint arXiv:1202.5822}, 2012.

\bibitem{PhysRevLett.123.070503}
E.~Campbell, ``Random compiler for fast hamiltonian simulation,'' {\em Phys.
  Rev. Lett.}, vol.~123, p.~070503, Aug 2019.

\bibitem{Chen_2021}
Y.-H. Chen, A.~Kalev, and I.~Hen, ``Quantum algorithm for time-dependent
  hamiltonian simulation by permutation expansion,'' {\em PRX Quantum}, vol.~2,
  Sept. 2021.

\bibitem{kim2023evidence}
Y.~Kim, A.~Eddins, S.~Anand, K.~X. Wei, E.~Van Den~Berg, S.~Rosenblatt,
  H.~Nayfeh, Y.~Wu, M.~Zaletel, K.~Temme, {\em et~al.}, ``Evidence for the
  utility of quantum computing before fault tolerance,'' {\em Nature},
  vol.~618, no.~7965, pp.~500--505, 2023.

\bibitem{King_2023}
A.~D. King, J.~Raymond, T.~Lanting, R.~Harris, A.~Zucca, F.~Altomare, A.~J.
  Berkley, K.~Boothby, S.~Ejtemaee, C.~Enderud, E.~Hoskinson, S.~Huang,
  E.~Ladizinsky, A.~J.~R. MacDonald, G.~Marsden, R.~Molavi, T.~Oh,
  G.~Poulin-Lamarre, M.~Reis, C.~Rich, Y.~Sato, N.~Tsai, M.~Volkmann, J.~D.
  Whittaker, J.~Yao, A.~W. Sandvik, and M.~H. Amin, ``Quantum critical dynamics
  in a 5,000-qubit programmable spin glass,'' {\em Nature}, vol.~617,
  pp.~61--66, apr 2023.

\bibitem{semeghini2021probing}
G.~Semeghini, H.~Levine, A.~Keesling, S.~Ebadi, T.~T. Wang, D.~Bluvstein,
  R.~Verresen, H.~Pichler, M.~Kalinowski, R.~Samajdar, {\em et~al.}, ``Probing
  topological spin liquids on a programmable quantum simulator,'' {\em
  Science}, vol.~374, no.~6572, pp.~1242--1247, 2021.

\bibitem{Preskill_2018}
J.~Preskill, ``Quantum computing in the {NISQ} era and beyond,'' {\em Quantum},
  vol.~2, p.~79, aug 2018.

\bibitem{Kadowaki_1998}
T.~Kadowaki and H.~Nishimori, ``Quantum annealing in the transverse ising
  model,'' {\em Physical Review E}, vol.~58, pp.~5355--5363, nov 1998.

\bibitem{farhi2000quantum}
E.~Farhi, J.~Goldstone, S.~Gutmann, and M.~Sipser, ``Quantum computation by
  adiabatic evolution,'' 2000.

\bibitem{JOHANSSON20121760}
J.~Johansson, P.~Nation, and F.~Nori, ``Qutip: An open-source python framework
  for the dynamics of open quantum systems,'' {\em Computer Physics
  Communications}, vol.~183, no.~8, pp.~1760--1772, 2012.

\bibitem{JOHANSSON20131234}
J.~Johansson, P.~Nation, and F.~Nori, ``Qutip 2: A python framework for the
  dynamics of open quantum systems,'' {\em Computer Physics Communications},
  vol.~184, no.~4, pp.~1234--1240, 2013.

\bibitem{Chen_2022}
H.~Chen and D.~A. Lidar, ``Hamiltonian open quantum system toolkit,'' {\em
  Communications Physics}, vol.~5, may 2022.

\bibitem{DifferentialEquations.jl-2017}
C.~Rackauckas and Q.~Nie, ``Differentialequations.jl – a performant and
  feature-rich ecosystem for solving differential equations in julia,'' {\em
  The Journal of Open Research Software}, vol.~5, no.~1, 2017.
\newblock Exported from https://app.dimensions.ai on 2019/05/05.

\bibitem{Lucas_2014}
A.~Lucas, ``Ising formulations of many {NP} problems,'' {\em Frontiers in
  Physics}, vol.~2, 2014.

\bibitem{Matsuda_2009}
Y.~Matsuda, H.~Nishimori, and H.~G. Katzgraber, ``Quantum annealing for
  problems with ground-state degeneracy,'' {\em Journal of Physics: Conference
  Series}, vol.~143, p.~012003, jan 2009.

\bibitem{magnus1954exponential}
W.~Magnus, ``On the exponential solution of differential equations for a linear
  operator,'' {\em Communications on pure and applied mathematics}, vol.~7,
  no.~4, pp.~649--673, 1954.

\bibitem{Blanes_2009}
S.~Blanes, F.~Casas, J.~Oteo, and J.~Ros, ``The magnus expansion and some of
  its applications,'' {\em Physics Reports}, vol.~470, pp.~151--238, jan 2009.

\bibitem{arnal2018general}
A.~Arnal, F.~Casas, and C.~Chiralt, ``A general formula for the magnus
  expansion in terms of iterated integrals of right-nested commutators,'' {\em
  Journal of Physics Communications}, vol.~2, no.~3, p.~035024, 2018.

\bibitem{An_2022}
D.~An, D.~Fang, and L.~Lin, ``Time-dependent hamiltonian simulation of highly
  oscillatory dynamics and superconvergence for schrödinger equation,'' {\em
  Quantum}, vol.~6, p.~690, Apr. 2022.

\bibitem{sanchez2011new}
S.~S{\'a}nchez, F.~Casas, and A.~Fern{\'a}ndez, ``New analytic approximations
  based on the magnus expansion,'' {\em Journal of mathematical chemistry},
  vol.~49, pp.~1741--1758, 2011.

\bibitem{bqpjson}
Y.~Pang and C.~Coffrin, ``Bqpjson.''
  \url{https://github.com/lanl-ansi/bqpjson}, 2019.

\bibitem{morrell2023signatures}
Z.~Morrell, M.~Vuffray, A.~Y. Lokhov, A.~B{\"a}rtschi, T.~Albash, and
  C.~Coffrin, ``Signatures of open and noisy quantum systems in single-qubit
  quantum annealing,'' {\em Physical Review Applied}, vol.~19, no.~3,
  p.~034053, 2023.

\bibitem{bloqade2023quera}
``Bloqade.jl: {P}ackage for the quantum computation and quantum simulation
  based on the neutral-atom architecture.,'' 2023.

\end{thebibliography}
LA-UR-24-28104

\end{document}